# Dissociating spatial frequency reliance from adversarial robustness advantages in neurally guided deep convolutional neural networks

Zhenan Shao[1,2,3 *], Tianyu Ren[4,5], Chengxiao Wang[5], Leyla Isik[3], Diane M. Beck[1,2]

[1] Department of Psychology, University of Illinois Urbana-Champaign

[2] The Beckman Institute, University of Illinois Urbana-Champaign

[3] Department of Cognitive Science, Johns Hopkins University

[4] Department of Civil and Environmental Engineering, University of Illinois Urbana-Champaign

[5] Department of Computer Science, University of Illinois Urbana-Champaign

* Corresponding author

E-mail: zshao19@jh.edu (Z.S.)

# Abstract

Deep convolutional neural networks (DCNNs) have approached and even surpassed human-level performance on many visual recognition tasks, yet they remain strikingly vulnerable to near-imperceptible perturbations generated by adversarial attacks. Recent research demonstrates that aligning DCNN representations with human visual cortex activity improves adversarial robustness, but the mechanisms driving this advantage are yet to be understood. One hypothesis suggests that neural alignment confers robustness by biasing models away from brittle high-frequency details and towards the low spatial frequencies (LSF). However, recent work indicates that human object recognition critically depends on a narrow, mid-frequency "human channel". Interestingly, this band was partially preserved in prior LSF-focused studies. In this work, we explicitly investigate whether a spectral bias towards the LSF or the human channel is the primary driver of the adversarial robustness observed in neurally aligned DCNNs. We first show that DCNNs aligned to higher-order regions of the human ventral visual stream systematically increase their reliance on both the LSF and the human channel. However, directly steering DCNNs towards these bands revealed a clear dissociation. Biasing models towards the human channel, either alone or together with the LSF, does not improve robustness and can even impair it. LSF bias produced some robustness gains, but such improvements are modest despite inducing much larger shifts in spatial-frequency reliance than the neurally-aligned DCNNs. Spatial-frequency-biased models overall show little, if any, increase in similarity to human neural representational geometry. Together, our results suggest that altered spatial-frequency reliance is likely an emergent property of learning more human-like representations rather than the primary mechanism by which neural alignment confers adversarial robustness, and motivate the need for future research examining representational properties beyond spatial-frequency biases.

# Introduction

Deep convolutional neural networks (DCNNs) have achieved remarkable success across a wide range of visual recognition tasks, approaching and even exceeding human-level performance on tasks from classic image classification [1] to more demanding segmentation [2] and even visual reasoning tasks [3]. However, despite these advances, DCNNs remain surprisingly vulnerable to even near-imperceptible perturbations generated by adversarial attacks [4,5], which can reliably induce catastrophic failures in DCNNs' performance. This fragility stands in sharp contrast to the robustness of human visual perception [6–8] and has motivated growing interest in identifying principles that could endow artificial systems with more human-like robustness [9–12]. Recently, an emerging line of work [12–15] has shown that aligning DCNN representations with neural activity from primate and human visual cortex can substantially improve robustness to adversarial image perturbations. These findings not only demonstrate neural representational alignment as a promising strategy but also suggest that biological visual systems develop representations that are more suitable for robust visual inference than those learned by conventionally trained DCNNs. However, the mechanisms by which neural alignment confers such adversarial robustness remain poorly understood. In particular, it is unclear which properties of biological visual representations are inherited by DNNs and how these properties directly support improvements in adversarial robustness. Clarifying these mechanisms is critical not only for advancing our understanding of the remarkable robustness of human vision, but also for guiding the development of more reliable artificial vision systems by isolating and replicating these principles in DCNNs.

One hypothesis that has emerged from recent work [11,16–20] is that neural alignment may improve robustness by biasing DCNNs towards different spectral features in natural images.

Spatial frequency (SF) is one of the most fundamental dimensions studied in human vision [21–23]. Different SF bands have long been proposed to convey distinct types of information in natural images. In particular, low spatial frequencies (LSF) convey coarse, global image structure that may provide a more robust basis for object recognition, whereas high spatial frequencies (HSF) encode fine details and local texture that can vary substantially under transformations [19,22,24]. Interestingly, the human ventral visual stream shows systematic changes in spatial frequency tuning with higher-order regions exhibiting increasing preference for LSF. The ventral stream, in particular, is a hierarchy of cortical regions extending from primary visual cortex (V1) to inferior temporal (IT) cortex that is critical for object recognition [25] and human visual robustness [7,9]. Furthermore, recent work [11] has also explicitly linked robustness gains from neural alignment to LSF reliance. For example, neurally aligned models have been reported to show increased reliance on LSF [11], and that explicitly biasing DCNNs towards LSF via architectural modifications or data augmentation can improve adversarial robustness [11,16,18]. These findings have thus led to the proposal that neural alignment enhances robustness primarily by steering DCNNs towards LSF features.

However, instead of implicating the LSF, recent work [26] has identified a much narrower frequency band, a one-octave-wide "human-channel" centered in the mid-frequency range, as critical for human object recognition. Human recognition performance was found to be severely impacted if the human channel is corrupted with noise, but perturbations outside this range, including the LSF, have little effects. Conventionally trained DCNNs, by contrast, tend to rely on a broader and more variable spectrum of spatial frequencies. Importantly, the degree to which a given DCNN concentrates its reliance on this human channel was also found to correlate with adversarial robustness [26], thus suggesting that robustness may depend not on a general shift towards LSF, but on selective reliance on this human channel. These findings are also consistent with a broader body of work demonstrating the importance of mid-frequency information for the recognition of letters [27,28], faces [29,30], and objects [31]. Interestingly, prior manipulations aimed at imposing an LSF bias also included information from the human channel [11,16]. In particular, Gaussian blur filters attenuate high spatial frequencies gradually and can preserve nontrivial energy within the mid-frequency human-channel range depending on the kernel parameters used (see Fig. 3a. for examples). As a result, adversarial robustness gains previously attributed to LSF bias could also be due to the preserved human-channel information. These observations thus raise a critical unresolved question: does neural alignment improve adversarial robustness by promoting reliance on the LSF, or by emphasizing information in the mid-frequency human channel, or both?

In the current work, we ask whether spectral bias towards the LSF range or the human channel can account for the adversarial robustness gains observed in neurally aligned DCNNs. We first characterize the spatial-frequency reliance profiles of a set of DCNNs aligned to neural representations from multiple regions along the human ventral visual stream [12], which allow us to systematically test whether robustness gains covary with reliance on specific frequency bands. We then use targeted manipulations to steer DCNNs towards the LSF range, the human channel, or both, and ask whether these manipulations reproduce the robustness gains and human-like representational geometry associated with neural alignment. We find that, in neurally aligned models, robustness is strongly associated with increased reliance on both the human channel and the LSF range. However, directly imposing these biases reveals a dissociation: human-channel bias does not improve robustness and can even impair it, whereas LSF bias produces only modest gains despite inducing large shifts in spatial-frequency reliance profiles. Moreover, spatial-frequency-biased models show little to no increase in similarity to human neural representations. These results suggest that altered reliance on the LSF range and the human channel is an emergent property of learning more human-like representations, rather than a sufficient explanation for the robustness benefits conferred by neural guidance.

# Results

## Neural guidance improves DCNNs' adversarial robustness

Adversarial attack (Fig. 1a) [4,5] has started to attract attention from the vision neuroscience community [10,12–14,17,32,33] as one of the most striking failures observed in artificial neural networks. The observation that tiny differences in inputs can alter model prediction suggests the representational geometry of models may be highly fragmented instead of smooth and differentiable as observed in primates [17,33]. These gradient-based attack algorithms also act as a precise probe to systematically identify the most sensitive features driving a model's predictions. The resulting catastrophic failures of DCNNs thus reveal their dependence on features that, while highly predictive of image labels, are brittle and poorly aligned with those used by human visual processing [34]. Consequently, achieving robustness to these perturbations requires fundamental changes that shift models towards more biologically aligned representational geometry and featural dimensions.

Concretely, adversarial attacks operate by constructing perturbations to model inputs, e.g. an image, that cause the model to make an incorrect prediction, given a controlled budget on perturbation magnitude. Formally, for each input image $X_i$, an adversarial perturbation $\tau$ could be obtained by maximizing the task loss while subjecting to a predefined maximum perturbation magnitude threshold, $\epsilon$, as measured by a specified $l_p$ norm:

$$\tau = \underset{||\tau||_p \leq \epsilon}{\operatorname{argmax}} \, \ell_p(f_\theta(x_i + \tau), y_i) \quad (1)$$

Where $f_\theta$ denotes the model and $y_i$ denotes the ground-truth label. As shown in Fig. 1b, such attacks can be effective even when perturbations are virtually imperceptible to the human eyes (Fig. 1b). Larger bounds $\epsilon$ generally lead to more powerful attacks, but even visually noticeable noise patterns do not fool humans under natural viewing conditions.

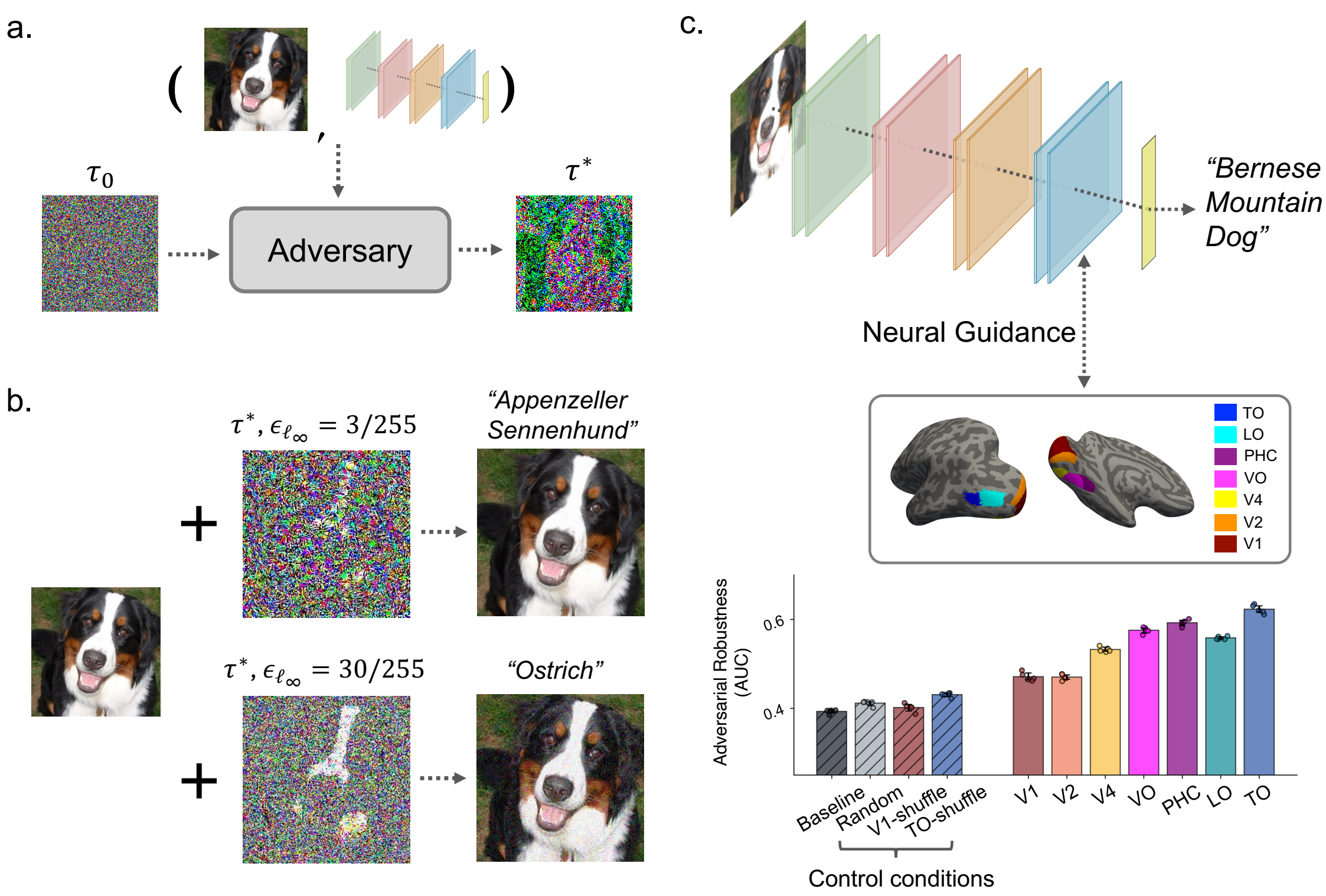

**Fig. 1 Overview of adversarial attacks, neural guidance, and adversarial robustness effects.** **a.** Schematic illustration of adversarial attacks. Given an input image and a trained model, an adversarial attack algorithm iteratively optimizes a bounded perturbation to produce an adversarial example $\tau^*$, which causes the model to misclassify an otherwise correctly recognized input. **b.** Example adversarial perturbations generated using an $\ell_\infty$-bounded projected gradient descent (PGD) attack. Perturbations are shown for a commonly used perturbation bound ($\epsilon = 3/255$; top), under which adversarial noise is virtually imperceptible to humans, and for an exaggerated bound ($\epsilon = 30/255$; bottom), included for visualization purposes only. **c.** Upper:

Illustration of the neural guidance framework [12]. During training, the penultimate layer of a deep convolutional neural network (DCNN) is encouraged to align with neural representations from a specific region of the human ventral visual stream, while simultaneously optimizing task performance. Bottom: Adversarial robustness, quantified as the area-under-the-curve (AUC) which was computed from the top-1 ImageNet classification accuracy under $\ell_\infty$-based PGD attacks across multiple $\epsilon$ values, increases systematically when guidance is applied using progressively higher-order ventral stream regions. Higher values indicate greater robustness. Dots in each condition indicates the performance from each of five independent training runs with different random seeds and the error bar denotes boot-strapped 95% confidence interval. Control models include a baseline DCNN without guidance, a random control that aligns DCNNs to unstructured noise signals, and V1- and TO-shuffle controls in which neural targets are mismatched across images, thereby disrupting meaningful representational alignment. This plot is adapted from [12] with permission.

Interestingly, a growing body of work [12,14,15] has demonstrated that such vulnerability could be alleviated when constraining or guiding DCNN representations to be more biologically grounded by explicitly aligning them to primates' neural representations. In particular, prior work [12] demonstrated in a framework named "Neural Guidance" (NG, Fig. 1c) that aligning DCNN representations to neural activity recorded in seven consecutive areas along the human ventral visual stream (VVS) leads to correspondingly hierarchical improvements in adversarial robustness. Specifically, DCNNs guided by higher order ventral stream regions, such as the Temporal-Occipital (TO) area, show greater resistance to adversarial attacks than models guided by earlier visual areas such as V1.

As shown in Fig. 1c, these neurally guided (NG) models also consistently outperform multiple control conditions, including conventionally trained models without neural guidance (“Baseline”), models guided by randomly generated and thus unstructured noise signals (“Random”), and shuffled neural alignment controls in which a model’s representation for a given image is aligned to neural responses evoked by different images (V1-shuffle and TO-shuffle). The observation that robustness improvements scale with the cortical source of guidance and are absent under random or shuffled controls suggests that these gains arise from information uniquely encoded in ventral stream representations, rather than from generic regularization effects [35] or noise injection [36]. One possible explanation for such improvements is that such alignment could be steering DCNNs towards “gold-standard” visual features relevant for human object recognition. In the following sections, we use the NG model set described above to test whether the adversarial robustness gains can be accounted for by changes in SF reliance, specifically bias towards the LSF versus the human channel.

## Neural guidance alters DCNNs’ spatial frequency reliance profiles

We first ask whether neural guidance systematically alters how models rely on information across spatial frequency channels. Specifically, we test whether NG-models show increased reliance on LSF, the human channel, or both. To probe models’ spatial frequency reliance, we followed prior work [11,20] and used adversarial perturbations generated by adversarial attacks as a diagnostic tool. Adversarial perturbations (Eq. 1) are optimized to maximally disrupt a model’s predictions with a limited budget. Consequently, these perturbations should direct their limited energy towards the input components to which the model is most sensitive. We therefore estimated each model’s SF reliance by Fourier-decomposing its successful adversarial perturbations, radially summing their power spectra across orientations, and normalizing by total perturbation power to obtain a one-dimensional SF reliance profile (see Methods for full details). In the resulting power spectra, higher power concentration in a given frequency band thus reflects greater model reliance on that SF range. Using this approach, we first computed the SF power spectra of adversarial perturbations generated with an $\ell_\infty$-projected gradient descent (PGD) attack [37] with $\epsilon = 4/255$ and compared frequency reliance profiles across all seven NG models (V1-, V2-, V4-, VO-, PHC-, LO-, and TO-guided) and four control models (Baseline, random, V1-shuffle, and TO-shuffle).

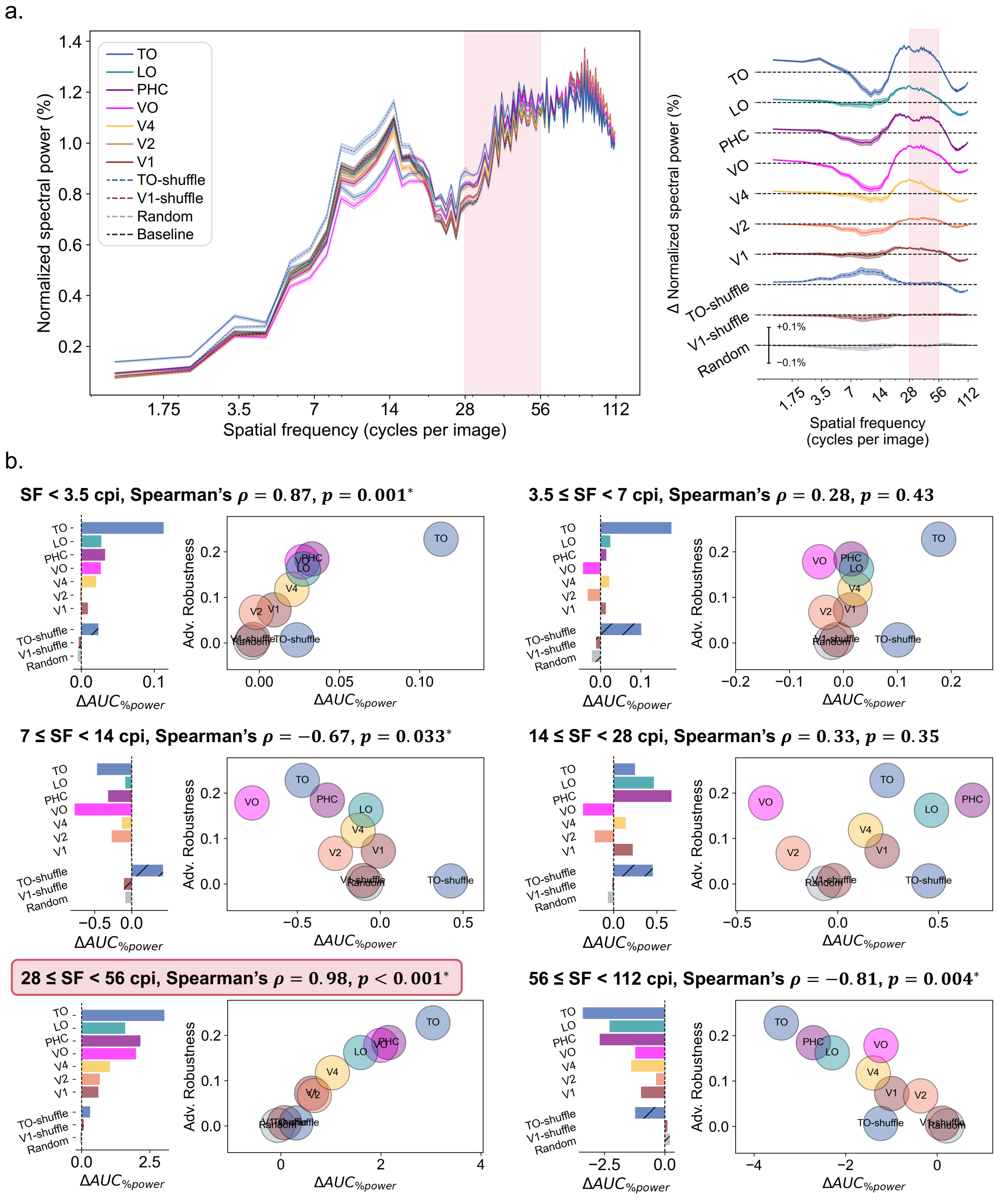

**Fig 2. Spatial frequency profiles of adversarial perturbations for NG and baseline models, and their correlation with robustness. a.** Left: Distribution of perturbation energy across spatial frequencies obtained from $\ell_\infty$-bounded PGD attacks for seven NG models and four

control models. Higher relative power in a given band indicates greater model reliance on that spatial frequency range. Shaded regions show 95% confidence intervals across adversarial perturbations generated for images in the evaluation set. The x-axis is plotted in octaves, with the human channel (28–56 cycles per image (cpi)) highlighted in pink. Right: Difference in spectral power between each model and the baseline model, revealing finer-grained variations in spatial frequency reliance profiles. Each curve represents a single model's deviation from baseline across the frequency spectrum. All curves share a common y-axis scale, indicated by the scale bar shown on the random model's trace. **b.** Octave-by-octave analysis. In each octave, left panels show the change in adversarial power allocation ($\Delta AUC_{\%power}$) for each model relative to the baseline model by calculating the area-under-curve (AUC). Right panels show correlations between adversarial power allocation ($\Delta AUC_{\%power}$) and adversarial robustness improvements across NG and control models shown in Fig. 1c. Spearman's $\rho$ and corresponding p-values are shown for each octave.

As shown in Fig. 2a, adversarial perturbation spectra showed a broad structure shared across all models, including the unguided baseline and control models. Perturbation power was relatively low at the lowest spatial frequencies and increased towards the high-frequency ranges. This common profile is consistent with prior observations that standard DCNNs can exploit fine-scale, high-frequency image components and that adversarial vulnerability is closely linked to frequency-domain sensitivity [19,20]. However, spatial frequency power concentration profiles differed systematically across models as a result of neural guidance. To quantify these differences, in Fig. 2b, we divided the frequency axis into octave bands and measured the proportion of spectral energy within each octave for every model. We then correlated these frequency-specific power allocations to adversarial robustness improvements shown in Fig. 1c using Spearman's $\rho$ to quantitatively assess the relationship between models' frequency reliance profiles and robustness. This analysis revealed several patterns.

First, within the human channel (28–56 cycles per image (cpi)), NG models guided by higher-order VVS regions showed greater reliance on this channel compared to the baseline models and models guided by earlier-stage brain regions. Importantly, the spectral power of perturbations in this channel correlated particularly strongly with robustness improvements (Spearman's $\rho = 0.98$, $p < 0.001$, Fig. 2b).

Second, NG models guided by higher-order VVS regions also showed greater reliance on the LSF range, as shown by the increased perturbation power in spatial frequencies below 3.5 cpi. This increased LSF reliance was also significantly correlated with adversarial robustness gains observed previously in NG models (Spearman's $\rho = 0.87$, $p = 0.001$, Figure 2b), although not as strong as the correlation in the human channel discussed above.

Third, all NG models showed a drop in reliance on the high spatial frequency (HSF) octave (56–112 cpi) relative to the unguided baseline model, as shown by the uniformly negative $\Delta AUC_{\%power}$ values (Fig. 2b). Moreover, across models, robustness was significantly anticorrelated with perturbation power concentration (Spearman's $\rho = -0.81$, $p = 0.004$) in this high-frequency octave, such that models that relied less on this octave tended to be more robust and were the ones guided by higher-order VVS regions. Together with the simultaneously increased reliance on LSF described above, these results indicate that neural guidance shifts models' reliance away from information conveyed in the HSF octave and towards that in LSF. Such results in these two octaves align well with prior observations [38] that conventionally trained DCNNs are biased towards fine-scale texture cues, while human object recognition rely almost entirely on objects' shape coded in the LSF range.

Interestingly, we also observed a significant negative correlation in the 7–14 cpi octave. In particular, reduced reliance on this octave was associated with greater robustness (Spearman's $\rho = -0.67$, $p = 0.033$; Fig. 2b), suggesting that the relationship between SF reliance and adversarial robustness may not be simply monotonic along a low-to-high frequency axis, cautioning against broad, binary LSF and HSF divisions. NG models' reduced reliance on this octave suggest that this intermediate range contains information that is neither preferred by higher-order brain regions nor advantageous for robustness.

Importantly, these observations were not specific to the particular attack configuration used above to generate perturbations for SF reliance analysis. We observed highly consistent results across different perturbation norms, magnitudes, and attack algorithms. Specifically, we repeated the analysis using two additional $\epsilon$ values ($\epsilon = 3/255$, $\epsilon = 5/255$), $\ell_2$-PGD attacks, and $\ell_\infty$-fast gradient sign methods (FGSM) attacks. All configurations yielded highly consistent spatial frequency reliance profiles (see Fig. S2 for full results).

Together, these results provide stronger correlational evidence for the hypothesis that neural guidance may confer robustness by steering models towards more human-relevant SF features, including both the human channel and the LSF range.

## Eliciting SF Bias in DCNNs towards the human channel or the LSF

Having shown that neural guidance systematically alters SF reliance in DCNNs, we next asked whether the adversarial-robustness gains of NG models can be directly attributed to increased reliance on human-relevant SF bands, specifically the human channel or the LSF.

Here, we test this hypothesis by explicitly biasing models towards these candidate SF bands during training and evaluating the subsequent adversarial robustness performance.

Motivated by prior work [16], we introduced SF bias through the data augmentation pipeline by replacing a subset of training images (30% or 50% depending on augmentation strength, see Methods for details) with SF-manipulated versions. To do so, we used a phase-scrambling manipulation [39] that preserved phase information within a targeted SF band while scrambling phase information outside that band. We included four conditions, each designed to target a distinct SF range (see Fig. 3a). The first condition ("human channel") targeted preserved the human channel using a bandpass mask whose parameters were derived from prior estimates of the SF range critical for human object recognition [26]. The second condition ("extreme-LSF") preserved only frequencies below 3.5 cpi, a range that does not overlap with the human channel but showed a significant positive correlation with robustness gains in NG models in our earlier analyses (Fig. 2b).

Unlike the human channel, whose bounds can be directly defined in cpi based on prior work [26], the precise LSF range most relevant to robustness is less well specified in the literature. We therefore included two additional LSF-biasing conditions whose preserved SF ranges were derived from filters used in previous studies. One condition ("fixed-$\sigma$-LSF") preserved SF ranges defined by a Gaussian blur kernel with a fixed standard deviation, $\sigma = 1.5$, a setting previously reported to yield the highest robustness benefits [11]. The last condition ("mixed-$\sigma$-LSF") used a mixture of Gaussian blur kernels with different $\sigma$ values that more faithfully reflect the human visual experience [16]. Importantly, images were not blurred in any of the conditions, rather, these filters were used to define the mask for phase-scrambling the untargeted SF bands and thus preserving on the targeted one (see Methods).

Notably, the SF ranges preserved in both the fixed-$\sigma$-LSF and mixed-$\sigma$-LSF conditions overlap with the human channel (Fig. 3a). Our design therefore yields clear and testable predictions. If the human channel is uniquely critical for robustness, then these two conditions should outperform both the extreme-LSF condition and the baseline model because they partially preserve the human channel. However, they should still trail the human-channel-only condition. On the other hand, if the LSF range is critical, then the human-channel-only condition should provide little or no benefit. If both factors contribute, then the fixed-σ-LSF and mixed-σ-LSF conditions, which preserve both the LSF range and part of the human channel, should show the greatest gains.

For each condition, we also varied the proportion of training images subjected to this phase-scrambling manipulation, thereby creating weak and strong SF biases. If a given SF manipulation is effective, stronger bias should be expected to amplify any corresponding robustness effect, thereby allowing a more graded interpretation of the results. Finally, we trained a matched set of baseline models using the same training recipe but without any imposed SF bias.

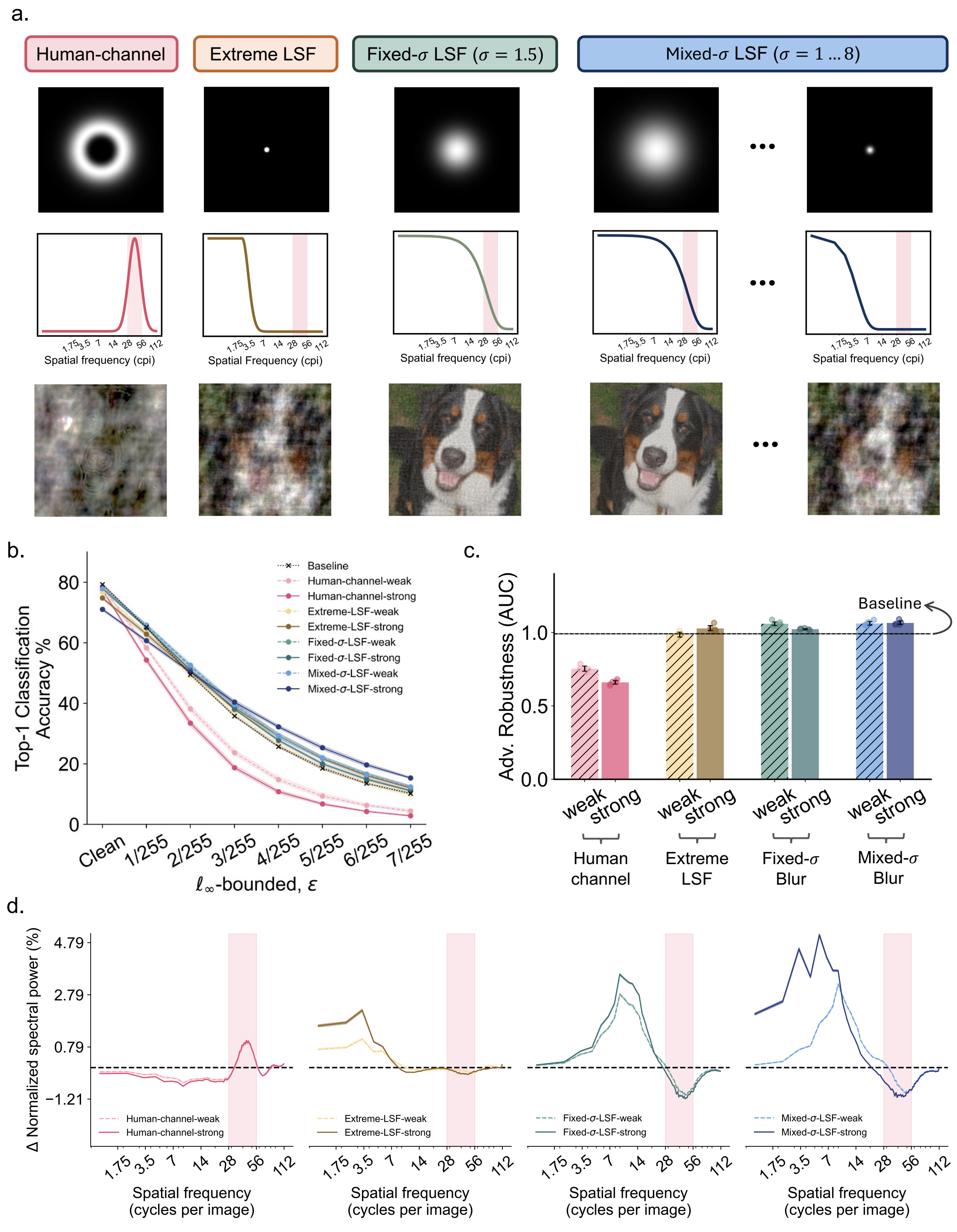

**Fig 3. Adversarial robustness and spatial frequency reliance profiles of DCNNs biased towards the low spatial frequency (LSF) range or the human channel using selective phase scrambling. a.** Visualization of the SF masks and example phase-scrambled images used to

impose SF biases during training. For each condition, we show the 2D frequency-domain mask that defines the SF range spared from phase scrambling (top), the corresponding 1D magnitude response plotted as a function of SF in cycles per image (cpi) on a logarithmic scale (middle), and example images in which all untargeted SF bands were phase scrambled while the targeted band was left intact (bottom). The four conditions are: 1. The human-channel bandpass condition with parameters derived from [26]; 2. an extreme low spatial frequency (LSF) low-pass condition preserving frequencies below 3.5 cpi, which showed a significant correlation with robustness gains in neurally guided models in Fig. 2; 3. a fixed-$\sigma$ LSF condition using a Gaussian kernel with standard deviation $\sigma = 1.5$, following [11], and 4. a mixed-$\sigma$-LSF condition implemented as a mixture of Gaussian kernels each with different $\sigma$s, following [16], for which two example kernels ($\sigma = 1.0$, $\sigma = 8.0$) are shown here. The human channel is highlighted by the pink shaded region in all 1D magnitude response plots to show that both the fixed- and mixed-$\sigma$ LSF conditions partially preserve the human channel. **b.** Adversarial robustness performance of models trained under each SF-bias condition. In all conditions except for baseline, we included weak and strong SF bias variants by manipulating the proportion of filtered images during training. Top-1 classification accuracy under $\ell_\infty$-based PGD adversarial attacks is shown across a range of perturbation bounds $\epsilon$. Shaded regions indicate bootstrapped 95% confidence intervals (CI) estimated from across five random seeds used to initialize the models. **c.** Area under the robustness curve (AUC) summarizing adversarial performance for each condition shown in b. with higher AUC indicating greater robustness. Error bars denote bootstrapped 95%-CI. **d.** SF reliance profiles of all SF-biased models, estimated using the same perturbation spectrum analysis as in Fig. 2a. From left to right, panels show the human-channel, extreme-LSF, fixed-$\sigma$-LSF, and mixed-$\sigma$-LSF conditions. Elevation relative to the dashed baseline indicates increased reliance on that SF range relative to the baseline model. Weak and strong biases are plotted together using different colors. The human channel is again highlighted by the pink shaded region.

We first evaluated the adversarial robustness of these SF-biased models using $\ell_\infty$-bounded PGD attacks [37] across a range of $\epsilon$ values and summarized performance using the area under the robustness curve (robust AUC). As shown in Fig. 3b and c, consistent with prior literature [11,16], all LSF-biased models showed either comparable or slightly improved adversarial robustness relative to the baseline model. Surprisingly, however, the human-channel-biased models showed a marked reduction in adversarial robustness. In other words, biasing models towards the human channel was not merely ineffective, but harmful, and stronger bias led to further degradation in performance. One possible concern is that this result reflects an ineffective SF manipulation rather than a genuine effect of emphasizing this channel. To address this possibility, we examined the SF reliance profiles of these SF-biased models, estimated using the same perturbation-spectrum analysis applied to the NG models (Fig. 2b). Fig. 3d confirms that our manipulations were indeed successful in that elevated reliance can be observed in the targeted SF range relative to baseline. Thus, biasing the model towards the human channel indeed worsened adversarial robustness.

However, one additional possibility is that the human channel may be beneficial when combined with an LSF bias. In other words, robustness improvements could, in principle, depend on a joint contribution of these two channels, given that the NG models show simultaneously increased reliance on both bands (Fig. 2b). These models' SF reliance profiles, at the same time, showed reduced reliance on the human channel, leaving open the possibility that jointly increasing reliance on both channels might be required. We therefore performed an additional "combined-channel" experiment in which models were biased towards both the LSF range and the human channel by exposing training images to either the extreme-LSF-preserved or the human-channel-preserved manipulation (see Methods). As shown in Fig. 4, this combined-channel condition again produced worse adversarial robustness than the baseline model. Thus, the detrimental effect associated with human-channel bias was not offset by the modest benefit of LSF bias. Therefore, bias towards the human channel, per se, likely cannot explain the adversarial-robustness benefits conferred by neural guidance.

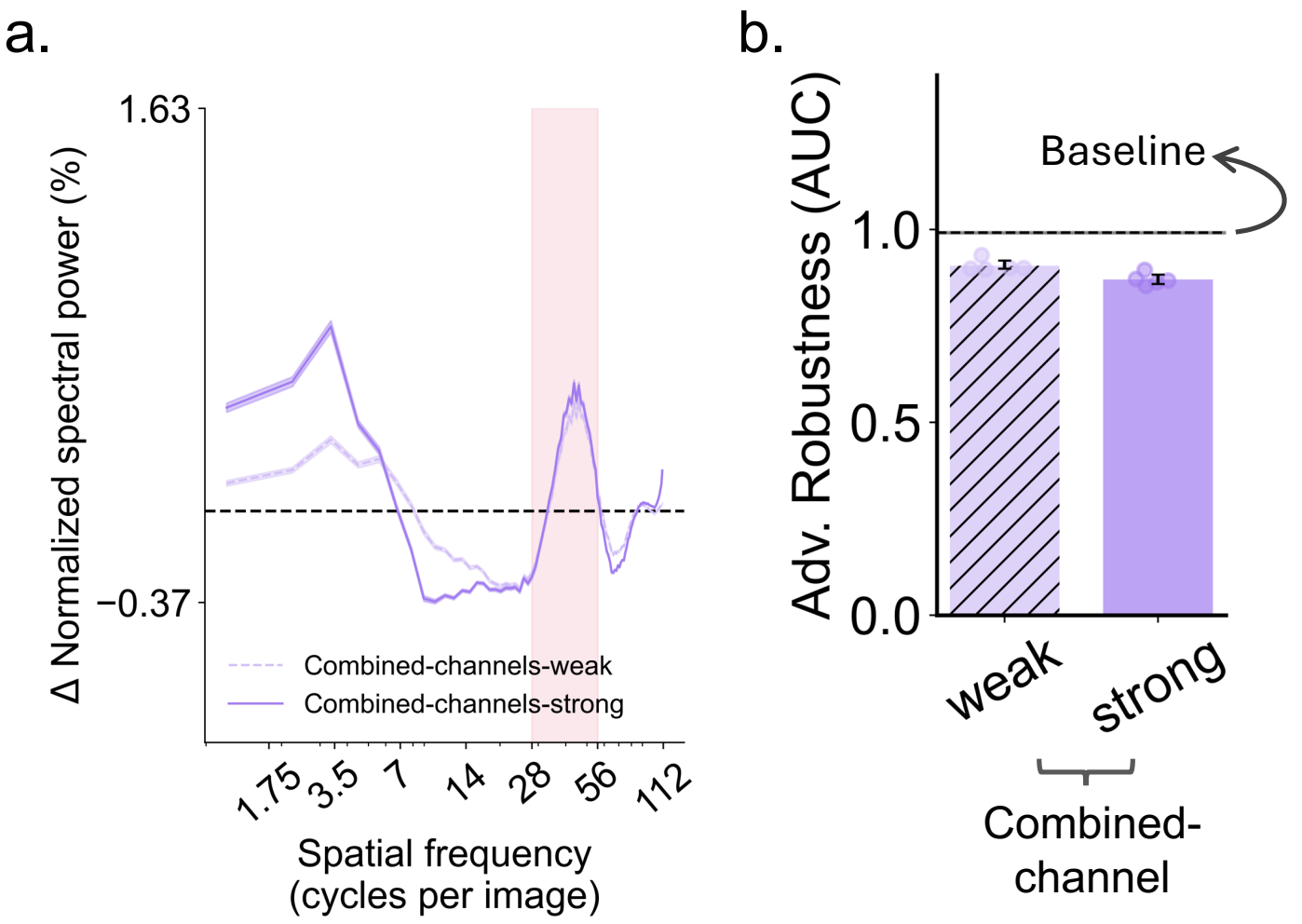

**Fig. 4. Spatial frequency reliance and adversarial robustness of models jointly biased towards the human channel and the extreme-LSF. a.** SF reliance profile of models trained in the combined-channel condition, in which the human channel and the extreme-LSF range (< 3.5 cpi) were spared while the remaining SF bands were phase scrambled. Both weak and strong bias variants are shown. The human channel is highlighted by the pink shaded region. **b**. Adversarial robustness of models trained under the combined-channel condition. Robustness is again summarized as the area under the robustness curve (AUC), with higher AUC indicating greater

robustness. The dashed horizontal line shows the mean performance of the baseline model, and the shaded region indicates its bootstrapped 95% CI. Dots denote individual models trained with five different random seeds, and error bars indicate bootstrapped 95% CI.

Could the modest robustness advantage of the LSF-biased models in Fig. 3c nevertheless indicate that LSF bias accounts for the benefits of neural guidance? This seems unlikely as the relationship between the induced SF shift and the resulting robustness gain is highly disproportionate. In particular, NG models show relatively small shifts in SF reliance (Fig. 2a) yet nontrivial robustness gains, whereas the explicitly LSF-biased models here show much larger SF reliance shifts but only modest gains. A direct comparison between the LSF-biased models and the NG models confirms this dissociation (Fig. S2). Therefore, although LSF bias may contribute to the benefits conferred by neural guidance, it is unlikely to be sufficient to account for the full robustness advantage observed in NG models.

In addition, we also evaluated the more commonly used filtering-based approach to manipulate the frequency contents of images [16,18,40,41]. However, as shown in Fig. S3, this method produced less spectrally precise shifts in SF reliance than phase scrambling. Filtered-image augmentation also yielded no robustness advantage even under LSF-biasing conditions, further reinforcing the conclusion that SF bias alone cannot fully account for the benefits observed in NG models.

## Limited necessity of the human channel for adversarial robustness

So far, explicitly biasing models towards the human channel has consistently harmed adversarial robustness. However, it remains possible that information within the human channel is needed to maintain robustness, given the critical role of human channel for human object recognition [26]. We therefore asked whether biasing models away from the human channel would be especially detrimental, relative to biasing models away from other SF channels such as the extreme-LSF range. To test this, we selectively phase-scrambled only the human-channel octave during training, thereby degrading information in that band, and compared model performance with both baseline models and models in which the extreme-LSF range (< 3.5 cpi) was selectively phase scrambled. If the human channel were uniquely necessary for maintaining adversarial robustness, then disrupting this band should produce the largest drop in performance.

Interestingly, as shown in Figure 5, selectively corrupting the human channel indeed reduced robustness, but the decline was smaller than that observed for the extreme-LSF condition. Therefore, although information in the human channel may contribute to adversarial robustness, it does not appear to be more necessary than the LSF range for maintaining it. Importantly, because the remainder of the spectrum was still available, models could in principle compensate by relying more heavily on other bands, including higher spatial frequencies. If part of the robustness loss reflects such compensatory reweighting, then the evidence for the human channel being uniquely necessary is even weaker. Overall, these results suggest that the human channel is, at most, only weakly necessary for adversarial robustness in DCNNs and cannot account for the benefits conferred by neural guidance.

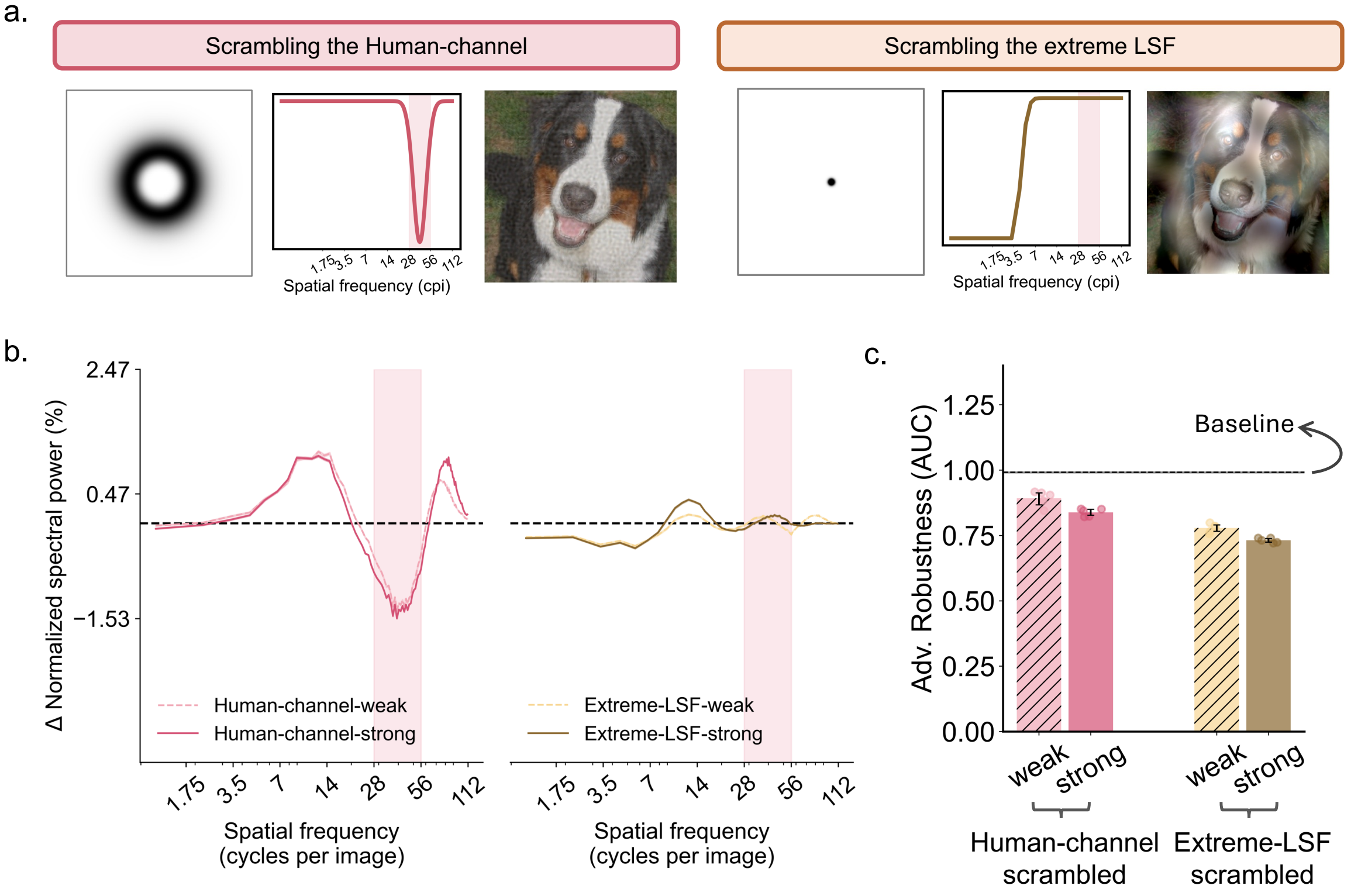

**Fig. 5 Selective phase scrambling of specific spatial frequency (SF) bands reveals limited necessity of the human channel for adversarial robustness. a.** Visualization of the SF masks and example phase-scrambled images used to bias models away from either the human channel or the extreme-LSF range. For each condition, we again show the 2D frequency-domain mask specifying the SF band selectively phase scrambled (left), the corresponding 1D magnitude response plotted as a function of SF in cycles per image (cpi) on a logarithmic scale (middle), and example images in which only the targeted SF band was phase scrambled while the remaining spectrum was left intact (right). The two conditions are: 1. selective scrambling of the human channel band-pass range; and 2. selective scrambling of the extreme-LSF range, preserving all other frequencies while corrupting frequencies below 3.5 cpi. **b.** SF reliance profiles of models trained under the two selective-scrambling conditions, estimated using the same perturbation-spectrum analysis as in Figs. 2a. The human channel is highlighted by the

pink shaded region. **c.** Adversarial robustness of these models, summarized as the area under the robustness curve (AUC) obtain in the same way as Fig. 3c. Higher AUCs indicate higher robustness levels. The dashed horizontal line indicates the baseline models' performance.

## Bias towards the human channel or the LSF does not lead to more human-like representational geometry

One possible resolution of the dissociation observed above is that the SF ranges relied on by NG models are not themselves critical for robustness, but rather happen to be where the robust features used by the brain are concentrated. In other words, SF biasing only teaches the model *where* in the Fourier domain to focus, but not *what* information to extract. SF-biased models may therefore continue to rely on classification-relevant features rather than those truly leveraged by the human visual system. NG models, by contrast, are directly trained by neural representations and thus have access to the information relevant for human vision. To test this, we next compare the representational geometry of SF-biased models and NG models to that of the human brain, hypothesizing that SF-biased models will fall short of matching human representational geometry.

In particular, we compared SF-biased models to three brain regions: the V1, V4, and the higher-order area TO ROIs, which has been shown to confer the largest robustness improvements when used to guide DCNNs among the regions tested in previous work [12]. These three regions approximately span early-, intermediate-, and higher-order stages of visual processing. Using classic representational similarity analysis (RSA) [42], we first constructed representational dissimilarity matrices (RDMs) for the three human ROIs by calculating pairwise Euclidean distances across fMRI response patterns for all images presented to participants in the NSD dataset [43]. We then computed model RDMs for all SF-biased models, including the human-channel, extreme-LSF, fixed-$\sigma$-LSF, mixed-$\sigma$-LSF, and combined-channel conditions, using representations elicited by the same image set. As a reference, we also constructed RDMs for NG models guided by each of these three regions, namely V1-guided, V4-guided, and TO-guided models [12]. These NG models provide a practical reference for the degree of representational alignment achievable under explicit neural supervision while maintaining task performance. Finally, we quantified representational similarity by calculating Spearman's $\rho$ between each model RDM and each brain-region RDM.

As illustrated in Fig. 6, the representational similarity of most SF-biased models to the brain remained close to that of the baseline models across all three brain regions, in contrast to the NG models. We did observe modest increases in similarity to V1 and V4 for the strongly LSF-biased mixed-$\sigma$ condition. However, these gains remained far below those observed in the NG models. For TO, we observed generally higher similarity scores across all model classes, including the baseline. This overall elevation may reflect stronger categorical clustering in higher-order regions [25], which more closely matches the models' classification objective. Critically, however, the SF-biased models still did not show a clear advantage over the baseline model. Notably, the combined-channel condition also failed to improve representational similarity relative to baseline, despite producing an SF reliance profile most similar to that of the NG models. Therefore, neither a bias towards the human channel nor a bias towards the LSF, alone or in combination, appears sufficient to bring DCNN representational geometry closer to that of the human brain to the extent observed in NG models, regardless of whether the reference region is V1, V4, or TO.

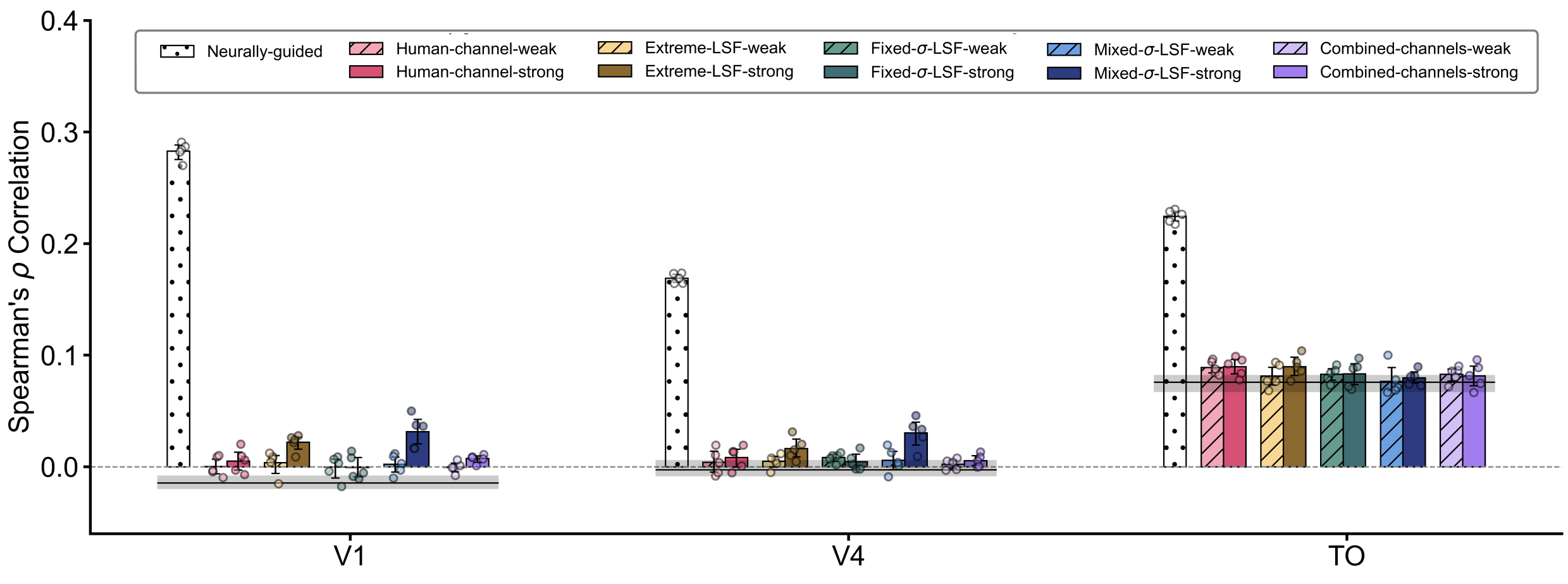

**Fig. 6: Representational similarity between spatial-frequency (SF)-biased models and human brain regions.** Representational similarity [42] between model representations and fMRI responses from V1, V4, and the temporal-occipital (TO) region, quantified as Spearman's $\rho$ between representational dissimilarity matrices (RDMs). White bars represent neurally guided models trained with direct supervision from the corresponding brain region and serve as a reference for the representational shift induced by explicit neural guidance. The horizontal black line denotes the representational similarity of the baseline model for each brain region. Distinct colors and hatching denote models trained under different SF-bias conditions. The grey shaded region surrounding the baseline line and the error bars on each bar indicate bootstrapped 95% confidence intervals calculated across five independent training seeds, whose individual values are also shown as dots.

# Discussion

We asked whether the adversarial robustness benefits conferred by neural guidance [12] could be explained by a shift in SF reliance towards bands identified to be critical for human vision, specifically the narrow mid-frequency “human channel” identified by Subramanian et al. [26] or the LSF emphasized in prior work [11,16]. We first characterized the SF reliance profiles of the neurally guided models which showed consistent and hierarchically increasing adversarial robustness when the representational guidance was from successive regions along the human ventral visual stream. These models showed increased reliance on the LSF, reduced reliance on the HSF, and, more strikingly, increased reliance on the human channel that was nearly perfectly correlated with robustness gains across neurally guided models.

However, despite these strong correlational relationships, explicitly imposing SF biases towards the human channel or the LSF revealed important dissociations. First, biasing models towards the human channel did not improve adversarial robustness and instead consistently impaired it. Moreover, when we explicitly disrupted the human channel, reduced robustness was observed but less so than disrupting the LSF, indicating that the human channel is not uniquely necessary for maintaining robustness. Critically, biasing models towards the human channel also failed to increase similarity to human neural representations. These findings therefore argue against the possibility that robustness gains previously attributed to LSF-based manipulations [11,16] were in fact driven by accidental preservation of the human channel, and suggest that increased reliance on the human channel is unlikely to be the mechanism underlying the benefits of neural guidance.

Biasing models towards the LSF, by contrast, produced modest robustness improvements and slight increases in representational similarity, broadly consistent with prior work [11,16,18,40]. However, these gains were disproportionately small relative to the large induced shifts in SF reliance. In other words, if increased LSF reliance were the primary mechanism by which neural guidance improves robustness, then imposing a much stronger LSF bias should have produced robustness gains and human-like representational geometry greater than, or at least comparable to, those observed in neurally guided models, but it did not. Therefore, increased reliance on the LSF may provide only a limited contribution. In addition, simultaneously biasing models towards both the LSF and the human channel, thereby creating a frequency-reliance profile shifts closer to that observed in the NG models, still worsened adversarial robustness and did not improve representational similarity. More broadly, therefore, SF bias alone cannot fully account for either the magnitude of the robustness gains or the shift towards more human-like representational geometry observed in NG models. Altered SF reliance therefore appears to be one consequence of learning more brain-like representations rather than the primary mechanism underlying the adversarial robustness benefits.

If reliance on the LSF or the human channel does not drive adversarial robustness, what does? One possibility is to move beyond feature-based mechanisms, which attribute robustness to preferential reliance on specific "gold-standard" visual features, and instead adopt a geometric account that focuses on how information is structured in representational space. For example, prior work [7,9,44] has hypothesized that robustness emerges from how the ventral visual stream transforms and organizes representations to develop increasingly linearly separable object manifolds, defined as the continuous surface formed by all identity-preserving transformations of an object in the neural state space. Critically, such geometric transformations do not require explicit coding of any particular "robust" feature. Instead, feature preferences can emerge as a consequence of optimizing for separability, or "untangling" of object manifolds: features that vary reliably between objects while remaining stable across identity-preserving transformations, such as the information carried in the LSF and the human channel or the coarse shape of the objects [38], naturally become emphasized. Conversely, features that vary unpredictably within and across objects, such as fine-grained high-frequency texture, become marginalized.

Indeed, prior work [12] has shown that neurally guided models develop more compact and more linearly separable category manifolds. Moreover, instilling only coarse geometric properties of human representational structure into DCNNs is already sufficient to qualitatively reproduce the adversarial robustness benefits. Our findings in the current work also fit into such a geometric framework of how the brain achieves visual robustness. Although the LSF and the human channel biases emerge reliably from neural guidance and correlate strongly with robustness, they could not fully explain either the full magnitude of the robustness gains or the accompanying shift towards human-like representational geometry. This pattern suggests that LSF or human-channel preference could also be an emergent correlate of a broader reconfiguration of representational geometry rather than the primary causal mechanism underlying adversarial robustness. Future work could therefore more directly characterize which and how specific geometric transformations are critical for conferring adversarial robustness.

One limitation of our approach is that, through data augmentation, we operationalized SF bias as a population-level property shaped by visual experience. In other words, the entire model was biased towards or away from specific frequency bands by manipulating the statistical properties of its training inputs. However, in the primate visual system, individual neurons show diverse tuning profiles even with a small cortical region [21,45,46]. Although the proportion of neurons preferring, for example, the LSF may increase along higher stages of the ventral stream [47], these regions do not exclusively encode any single channel. Moreover, these tuning properties may reflect fundamental architectural constraints of the visual system, such as from the receptive field structure [48,49], rather than learned preferences that can be flexibly reshaped by experience. Therefore, an alternative explanation for the absence of robustness benefits in our current work could be that architectural constraints within the network itself are needed to faithfully capture how SF tuning is instantiated in biological systems. Consistent with this possibility, Li and colleagues [11] reported that prepending a fixed, differentiable low-pass filtering layer to bias the model towards the LSF can improve adversarial robustness. However, such preprocessing layers also alter the gradients that drive not only learning but also white-box attacks, making it difficult to disentangle biologically meaningful effects of SF tuning from changes in optimization dynamics or gradient smoothing [55,56]. Future work could address this challenge by isolating SF tuning effects from these artifacts, and work towards more biologically-aligned solutions, such as constraining the tuning profiles of individual neurons to more faithfully mimic the properties of the primate visual cortex [48].

Notably, a growing body of work has specifically examined adversarial robustness through a spatial-frequency perspective. Much of this literature has framed the problem in terms of broad low-versus-high frequency trade-offs. Standard DCNNs can exploit high-frequency image components and fine-scale detail, whereas more robust models often show reduced sensitivity to such information and relatively greater reliance on lower spatial frequencies [11,19,40]. At the same time, the broader picture seems to be more nuanced. Frequency-dependent robustness trade-offs vary across corruption types [50], the impact of spatial-frequency constraints can depend strongly on the dataset [51], and successful adversarial perturbations can also be found in low-frequency or otherwise dataset-dependent subspaces [52,53]. Importantly, however, the aim of the current work is not to resolve the full debate over the spectral basis of adversarial vulnerability. Rather, we ask a more targeted mechanistic question: what properties of neural representations enable them to confer adversarial robustness? Specifically, we tested whether the robustness benefit of neural alignment arises from biasing models towards the LSF range, as previous work has suggested [11], or towards the more recently proposed human channel [26]. Our results present a clear answer for the latter and a qualified answer for the former. Preferential reliance on the human channel does not explain the robustness benefits of neural guidance, whereas LSF bias may provide a limited contribution but cannot reproduce either the magnitude of the robustness gains or the shift toward more human-like representational geometry observed in NG models. SF bias therefore seems best understood as one partial property of brain-like representations rather than the primary computational mechanism by which the human visual system achieves robustness.

# Materials and Methods

## Spatial frequency reliance profile of neurally-guided DCNNs

We examined the spatial frequency (SF) reliance profiles of seven neurally-guided (NG) models (V1, V2, V4, VO, PHC, LO, and TO) and four control models (Baseline, Random, V1-shuffle, and TO-shuffle) to test whether neural guidance systematically shifts reliance on specific SF channels, in particular, the LSF and the human channel, that predicts the previously observed adversarial robustness gains. More training details on these NG models can be found in [12]. Similar to prior work [11,20], we used adversarial perturbations as diagnostic probes of models' feature reliance. Gradient-based adversarial attacks identify directions in input space that most effectively alter model decisions. The spectral content of these perturbations, therefore, could be leveraged to reverse-engineer the spatial frequencies most critical to the model's classification behavior.

### Adversarial perturbation generation

For each input image $x_i \in \mathbb{R}^{H \times W \times 3}$ with ground-truth label $y_i$, we generated an adversarial perturbation $\tau$ under an $\ell_\infty$-norm constraint:

$$||\tau||_\infty \leq \epsilon, \quad (2)$$

with $\epsilon = 4/255$. This value is commonly used in ImageNet-based robustness benchmarks [54] and corresponds to a perturbation magnitude at which robustness differences between the NG and the control models are prominent [12]. However, we have replicated our results using other $\epsilon$ values including 3/255 and 5/255, both of which yielded near-identical SF reliance profiles (Fig. S1a).

Perturbations were generated using projected gradient descent (PGD) [37] adversarial attack, which approximately aims to solve eq. 1, although we have also replicated results with the fast gradient sign method (FGSM) [55] as shown in Fig. S1b. Specifically, PGD were initialized with uniform random noise $\tau_0 \sim U(-\epsilon, \epsilon)$, and iteratively updated it using gradient ascent with projection back onto the $\ell_\infty$-ball after each step, although bounding with $\ell_2$-ball yields similar results (Fig. S1c). Importantly, only perturbations that successfully changed the model's originally correct prediction were retained for subsequent spectral analysis to ensure that the perturbations reflect decision-relevant feature sensitivity of the models.

## Spectral analysis of adversarial perturbations

To quantify spectral properties of the adversarial perturbations, we first computed the two-dimensional discrete Fourier transform of each RGB channel of a given perturbation $\tau^*$. Let $\tau_c^*$ denote the perturbation for channel $c \in \{R, G, B\}$, the power spectrum $P(u, v)$ of $\tau^*$ is:

$$P(u, v) = \frac{1}{3} \sum_{c} |\mathcal{F}(\tau_c^*)(u, v)|^2 \tag{3}$$

where $(u, v)$ denotes the spatial frequency coordinates and $\mathcal{F}$ denotes DFT. To obtain the one-dimensional power distribution spectrum, we rotationally averaged Fourier coefficients over frequency bins defined by radial distance: $r = \sqrt{u^2 + v^2}$, and normalized the spectrum by its total power to yield a proportion-of-energy distribution.

We repeated this procedure for all successful adversarial perturbations in the test set of the curated ImageNet-50 subset (1944.55 $\pm$ 178.94 images were used for estimation on average across all 11 models, excluding the Baseline model as the adversarial robustness improvement and the SF reliance shifts were quantified with respect to the Baseline model). This particular subset was constructed to better match the image statistics of the MS-COCO stimuli used in the Natural Scenes Dataset [43], from which the neural guidance signals were derived to train the NG models. Details of this curation procedure can be found in Shao et al. 2025. Furthermore, for each model, we averaged across all perturbations spectrums to obtain the mean SF reliance profile. The 95% confidence interval were estimated via bootstrapping across perturbation spectrums with 10,000 samples. In addition, since robustness performance across five independent training seeds showed minimal variance (Fig. 1c) thus indicating high stability of model behavior, we conducted all spectral analysis on one instance for each model trained with random seed zero.

To link each model's SF reliance profile to its adversarial robustness, we partitioned the spatial frequency axis into octave bands defined in cycles per image (cpi). We treated the human channel identified in [26], spanning the one-octave wide interval $f \in [28, 56)$ cpi as one complete octave. We then defined frequency bands as successive intervals while maintaining octave scaling across the spectrum. Additionally, due to the relatively sparse sampling of Fourier coefficients at the lower end of spatial frequencies, we merged the two lowest octave intervals: $f \in (0, 1.75)$ and $f \in [1.75, 3.5)$ cpi, into a single extreme low spatial frequency band (extreme-LSF). Note that the DC component was removed. The resulting bands used in analysis were therefore: $(0, 3.5)$, $[3.5, 7)$, $[7, 14)$, $[14, 28)$, $[28, 56)$, $[56, 112)$, where 112 cpi represents the highest resolvable spatial frequency given the 224 × 224 ImageNet-standard image resolution and the Nyquist sampling constraint. Within each band, we quantified the shift in spectral allocation relative to the unguided Baseline model by computing the difference in area under the SF reliance spectrum ($\Delta AUC_{\%power}$), thereby measuring changes in power allocation attributable to neural guidance or the control manipulations. We then computed Spearman's $\rho$ rank correlation between these band-specific changes $\Delta AUC_{\%power}$, and the corresponding improvements in adversarial robustness across the NG models to test whether changes in particular SF bands predict robustness gains.

The adversarial robustness improvements of the NG models were evaluated using the same procedure applied to the SF-biased models in the current work described below. Full implementation details can also be found in [12]. Briefly, each model was evaluated under the $\ell_\infty$-bounded PGD attacks [37] across a range of perturbation magnitudes $\epsilon$. For each $\epsilon$, we computed the top-1 classification accuracy on the attacked test set, yielding a robust accuracy curve as a function of $\epsilon$. Robustness was then summarized as the area under the accuracy curve (robust AUC) and the difference in robust AUC between each model and the unguided Baseline model were used here to quantify improvement. Positive values thus indicate increased robustness relative to the Baseline.

## Training with spatial frequency bias

To bias DCNNs towards specific spatial frequency ranges, including the LSF and the human channel identified in [26], we introduced frequency-biasing augmentations into the standard ImageNet training pipeline. In addition to standard ImageNet augmentations including random cropping, resizing, and horizontal flipping, we applied band-selective phase scrambling to a subset of training images. In particular, we defined frequency-domain masks to determine which SF ranges retained their original phase structure and which ranges underwent phase scrambling, while the original Fourier amplitude spectrum was preserved. Importantly, evaluation images were never modified, ensuring that model performance was assessed on natural, unaltered inputs. We detail these conditions below.

## Phase-scramble-based spatial frequency biasing

### Biasing towards the human-channel

To bias DCNNs towards the narrow spatial frequency band reported to be critical for human object recognition [27], we first constructed the frequency-domain mask to determine which spatial frequency ranges retained their original phase structure and which ranges were phase scrambled. For the human-channel condition, we built a radially symmetric bandpass mask in the Fourier domain. Let $f$ denote spatial frequency in cycles per image (cpi), and $\log_2 f$ denote frequency expressed in octaves. The phase-preserving mask $M_{human}$ was then defined as a radially symmetric Gaussian envelope in log-frequency space as follows:

$$M_{human}(f) \propto \exp\left(-\frac{(\log_2 f - \mu)^2}{2\sigma^2}\right) \quad (4)$$

where $\mu$ specifies the center frequency in octaves, and $\sigma$ determines the bandwidth. These parameters were set according to estimates reported in [26], with $\mu = 4.5$, and $\sigma = 0.4246$. The mask was subsequently normalized to have unit peak amplitude. Using this mask, frequencies within the human channel will retain their original phase structure, whereas frequencies outside this range were phase scrambled.

Phase-scrambling is a commonly used manipulation for disrupting image structure while preserving the amplitude spectrum [39]. However, here, we extended the standard formulation to a soft-blended, frequency-selective variant that permits graded weighting across the spectrum. In particular, let $X_c(u,v) = A_c(u,v)e^{i\phi_c(u,v)}$ denote the Fourier transform of a given color channel $c$ of an input image $x$, where $A_c(u,v)$ and $\phi_c(u,v)$ are the amplitude and phase at a frequency coordinate $(u,v)$ respectively. For each image, we sampled a random phase field $\phi_r(u,v) \sim U(-\pi,\pi)$, shared across all three RGB color channels to preserve color coherence. We then constructed a new phase spectrum by combining the original and random phases in the complex plane, weighted by a soft mask $M(u, v)$:

$$\phi_c'(u, v) = \angle\left(M(u, v) \cdot e^{i\phi_c(u,v)} + (1 - M(u, v)) \cdot e^{i\phi_r(u,v)}\right) \tag{5}$$

where $M(u, v) = 1$ preserved the original phase, $M(u, v) = 0$ yields full phase scrambling, and intermediate values produce partial scrambling. The output image was then reconstructed use the original amplitude spectrum:

$$x_{i,c}' = \mathcal{F}^{-1}\left(A_c(u, v) e^{i\phi_c'(u,v)}\right) \tag{6}$$

where $\mathcal{F}^{-1}$ denotes the inverse Fourier transform. The original amplitude spectrum was retained exactly, and thus this manipulation preserved the overall spectral energy of the original image. Any residual out-of-range values after reconstruction were softly clipped before the standard ImageNet normalization step. Visualization of actual images used in the training can be found in Fig. 3a.

Finally, we controlled the strength of the SF bias by varying the probability $p$ that a training image was replaced by its selectively phase-scrambled version. Specifically, $p = 0.3$ was used to impose a weak bias and $p = 0.5$ to impose a strong bias towards the human channel.

**Biasing towards the low spatial frequencies (LSF)**

To bias models towards LSF information, we implemented two LSF-preserving phase-scrambling conditions. The first condition, termed “extreme-LSF”, targeted the extreme low spatial frequency range defined as the range $f < 3.5$ cpi, corresponding to the lowest two-octave ban, a range in which increased SF reliance was observed and such increment significantly predicted robustness improvements in the NG models (Fig. 2a). The corresponding phase-preserving mask was a radially symmetric Gaussian in log-frequency space, centered at the geometric mean of 1.75 and 3.5 cpi, i.e., $\mu = \sqrt{1.75 \times 3.5}$. Second, we additionally imposed a hard cutoff constraint such that SF below $\mu$ was also retained. Together, this produced a smooth, yet well-defined extreme-LSF mask preserving phase structure in the lowest two octaves while selectively scrambling phase outside that range.

Although this extreme-LSF range was identified in the spectral analysis on NG models, restricting intact phase information to such a narrow low-frequency range removes substantial image structure, as illustrated in Fig. 3a. Therefore, we introduced a second LSF-biasing condition, termed “fixed-$\sigma$-LSF”, using a more permissive low-pass mask based on Gaussian spatial smoothing with parameters identical to those used in [11]. Specifically, we derived the mask from the magnitude response of an isotropic Gaussian kernel with standard deviation $\sigma = 1.5$ to define a soft and more generous LSF-biasing mask. Importantly, this condition partially preserved mid-frequency content, including portions of the human channel, as shown in Fig. 3a. In contrast, the extreme-LSF condition did not overlap with the human channel.

Similar to the human-channel condition, we also controlled the strength of the SF bias by varying the probability $p$ that a training image was replaced by its selectively phase-scrambled version. Specifically, $p = 0.3$ was used to impose a weak bias and $p = 0.5$ to impose a strong bias towards the LSF.

**Biasing towards the LSF with a mixture of masks**

We additionally implemented a third LSF-biasing condition designed to approximate a more developmentally plausible LSF preference as proposed by Jang and Tong [16]. This condition, named “mixed-$\sigma$-LSF”, extended the single-mask Gaussian-based procedure above by using a mixture of phase-preserving masks derived from Gaussian kernels with different widths.

In particular, each training image was assigned a mask whose effective width was sampled from a predefined discrete distribution. All kernel widths and sampling probabilities were chosen to exactly replicate the protocol from Jang and Tong [16]. In the weak condition, $\sigma \in \{1, 2, 3, 4, 5\}$ with corresponding probabilities $\{0.2129, 0.0653, 0.0200, 0.0062, 0.0019\}$, respectively. In the strong mixed-$\sigma$-LSF condition, $\sigma \in \{1, 2, 4, 8\}$, each sampled with equal probability $p = 0.2$. Critically, each image received a single mask per training step; the mixture applies at the level of the training distribution, not within an individual image. A representative visualization of two of the Gaussian kernels used with $\sigma = 1$ and $\sigma = 8$ and its effect on images are shown in Fig. 3a. It is worth noting that, similar to the fixed-$\sigma$-LSF condition, this mixture-based mask partially preserves mid-frequency content, including portions of the human channel.

**Biasing towards both the human channel and the LSF**

We additionally implemented a "combined-channel" condition to bias models towards both the human channel and the extreme-LSF ($f < 3.5$ cpi) range. The same band-selective phase-scrambling procedure described above was used, with one difference in how masks were assigned. For each selected training image, we sampled with equal probability either the human-channel-preserving mask or the extreme-LSF-preserving mask, and applied only that single mask. The two manipulations were thus combined at the level of the training distribution rather than within individual images. This design successfully induced simultaneously increased reliance on these two ranges (Fig. 4a).

**Inverse-mask phase-scrambling for necessity-check**

We conducted necessity-check experiments to test whether selectively disrupting the human channel was especially detrimental to adversarial robustness compared to the LSF range. To accomplish this, we inverted the corresponding masks ($M_{inv} = 1 - M$) so that phase scrambling was concentrated within the SF band of interest while phase structure outside that band was preserved. In other words, phase structure in the targeted band, such as the human channels, is now disrupted. All other training procedures remained identical to previous conditions. Visualization of the masks and example phase-scrambled images can be found in Fig. 5a.

## Dataset, models and training details

All models were trained and evaluated on a classic image classification task using a 50-category subset of ImageNet (ImageNet-50) curated in [12]. This subset was originally constructed to enable direct comparison between model representations and human neural representations derived from the Natural Scenes Dataset [43], which were also used in our representational-alignment-to-human analyses (Fig. 6) critical for the current work. We retained the original ImageNet training and validation splits and standard ImageNet data augmentations including random resized cropping and horizontal flipping. Importantly, despite the use of grey-scale image in previous work [11,16,26], we here used full-RGB color images.

All experiments used ResNet-18 [56] as the backbone architecture. No architectural modifications were introduced, except that the final fully connected classification layer was replaced with a 50-way output layer corresponding to the ImageNet-50 categories. Frequency bias was introduced exclusively through modifications to the data augmentation pipeline as described in the preceding sections.

All models were trained from scratch without pretrained weights. Training was conducted for 60 epochs using stochastic gradient descent (SGD) optimizer with momentum 0.9 and weight decay of $1 \times 10^{-4}$. The initial learning rate was set to 0.1 and decayed by a factor of 0.1 at epoch 45. The batch size was 1024. Label smoothing of 0.1 was applied during training. All training hyperparameters and optimization procedure remained identical across all conditions.

In addition, for each experimental condition, including all frequency bias manipulation and bias strength level, we trained five independent model instances using different random seeds for weight initialization. All reported model performance metrics, unless otherwise specified, are presented as mean values with 95% confidence intervals estimated via bootstrapping with 10,000 resamples across the five seeds. All training used a single NVIDIA RTX 5090 GPU.

## Adversarial robustness evaluation

Adversarial robustness of all SF-biased models was evaluated using the common $\ell_\infty$-bounded PGD attacks to consistent with prior work [12]. Perturbation magnitudes were tested over seven thresholds: $\epsilon \in \{1/255, 2/255, 3/255, 4/255, 5/255, 6/255, 7/255\}$. For each model instance, adversarial examples were generated on the ImageNet-50 validation set (2500 images in total) independently at each $\epsilon$, and top-1 classification accuracy was computed under attack. To summarize robustness performance across perturbation strengths, we computed the area under the robustness curve (AUC), providing a single scalar summary of performance for each model.

Notably, for the neurally guided models, we showed robustness improvements relative to the unguided Baseline model, expressed as differences in robust AUC (Fig. 1c), and these relative AUC values were used in subsequent spectral analyses (Fig. 2) to be consistent with the prior work [12]. In addition, all SF biased models showed robustness that was comparable to or lower than the Baseline model, so plotting relative differences would have exaggerated small numerical deviations and potentially obscured the overall performance pattern.

## Assessing representational similarity of all models to humans

To evaluate whether SF bias induces more human-like representational geometry, we performed the classic representational similarity analysis [42]. Neural data were obtained from Subject 1 of the NSD dataset [43], using the 1,000-image evaluation split consisting of diverse MS-COCO natural image stimuli shown to human subjects. We analyzed responses from three ventral visual stream regions: V1, V4, and TO. The same set of 1,000 NSD images was also presented to all SF-biased models (both filter-based and noise-overlay-based methods, see Fig. 6), and internal representations were extracted for direct comparison with neural data. Specifically, image representations were taken from the penultimate layer of the models, i.e., the 512-dimensional feature vector produced by the average pooling layer immediately preceding the final 50-way fully connected classification layer. In addition to the SF-biased models, we included the corresponding neurally guided (NG) models as reference points to indicate the degree of similarity achievable under explicit neural alignment. For the human data, voxel-wise activation patterns within each ROI were used as image representations, with voxel responses flattened into a single vector per image.

For each model and each ROI, we constructed a representational dissimilarity matrix (RDM) by computing all pairwise Euclidean distances between image representations, yielding a $1000 \times 1000$ symmetric matrix. These RDMs characterize the representational geometry of each model or neural region. We then computed Spearman's $\rho$ correlation between the vectorized upper triangles of the RDMs as a second-order representational similarity measure. RSA was performed independently for each of the five randomly initialized instances per model condition. Reported values in Fig. 6 reflect the mean across seeds, with 95% confidence intervals estimated via bootstrapping with 10,000 samples over all model instances.

## Acknowledgements

This work was supported by NIH R01MH132826 awarded to L.I.

## Data and code availability statement

All code necessary to reproduce our experiments and results are publicly available here: https://github.com/shaox192/spatial-frequency-bias.git.

# Supplemental Information

## Fig. S1

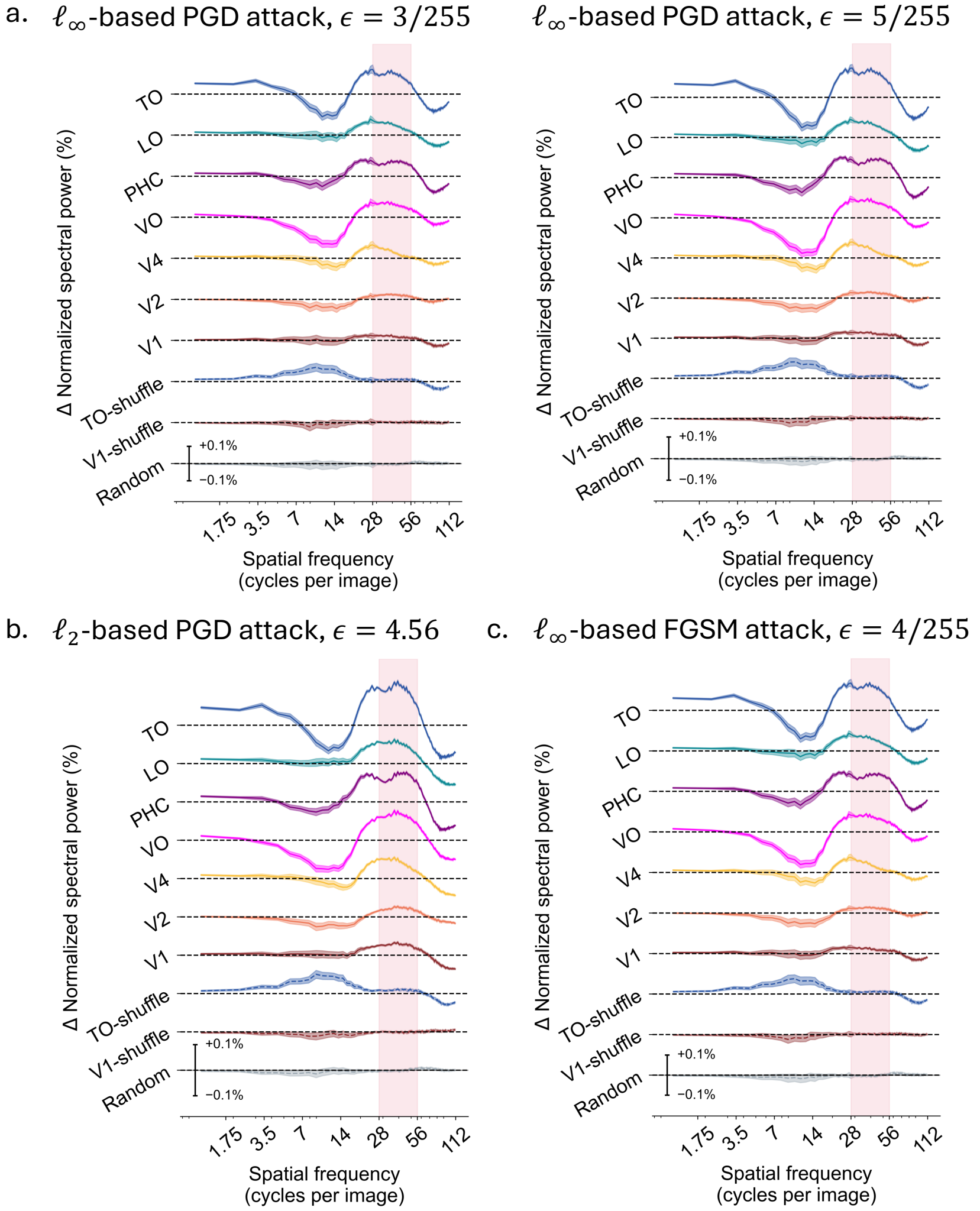

**Spatial frequency reliance profiles, shown as differences relative to the baseline model, for NG and control models estimated under alternative adversarial attack parameters.** Profiles shown are each computed using a. different perturbation magnitudes $\epsilon$ , b. an $\ell_2$-bounded attack instead of $\ell_\infty$, and c. the Fast Gradient Sign Method (FGSM). Across all conditions, the resulting profiles closely match those shown in Fig. 2a (right panel), which were obtained using an $\ell_\infty$-

bounded PGD attack with $\epsilon = 4/255$. These results indicate that the observed spatial frequency reliance patterns are agnostic to the choice of adversarial attack method and parameterization.

# Fig. S2

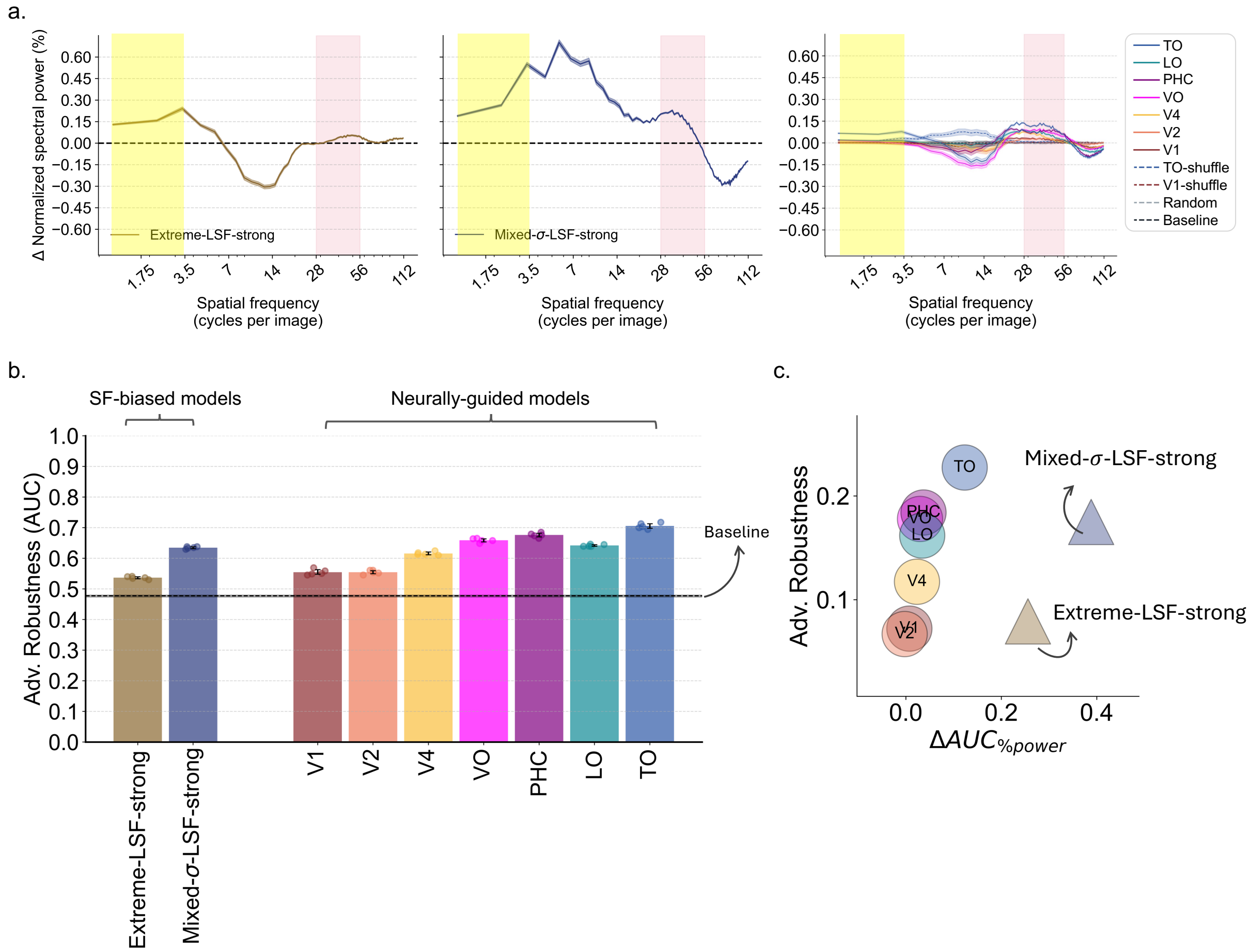

**Comparing the magnitude of SF reliance shifts and adversarial robustness gains in LSF-biased and neurally guided models.** To directly compare LSF-biased models with the neurally guided (NG) models, we trained a separate set of LSF-biased models maximally matching the training recipe used in the previous study [12]. This matched comparison was necessary because the LSF-biased models analyzed in the main text were trained under different conditions. Most notably, as described in the Methods, all models throughout the current work, except this particular set, are trained from scratch rather than from pretrained weights as the NG models do. We adopted from-scratch training throughout because the SF biases inherited from pretrained

weights were difficult to override, even with full-model fine-tuning. Consistent with this, the LSF-biased models shown here exhibit smaller shifts in SF reliance than the models shown in the main text (Fig. 3d).

Here, we focused on the two LSF-biasing conditions that produced the largest robustness gains in Fig. 3c, namely the strongly biased extreme-LSF model and the strongly biased mixed-$\sigma$-LSF model. For the extreme-LSF condition, we increased the proportion of filtered images needed to 80% (as opposed to 50% used in the main text) to induce a sufficiently strong bias, while keeping the filtering parameters otherwise matched to the main-text setting.

**a.** We first show the SF reliance profiles for the strongly biased extreme-LSF model, the strongly biased mixed-$\sigma$-LSF model, and the NG models. The human channel and the LSF range are highlighted in pink and yellow, respectively. Notably, both the LSF-biased models show substantially larger shifts in SF reliance than the NG models. The extreme-LSF model shows approximately a two-fold larger shift, whereas the mixed-sigma LSF model shows shifts of up to four-fold larger.

**b.** Adversarial robustness, summarized as robust AUC, for the two LSF-biased models and the seven NG models. Dots indicate performance of models trained using five different random seeds, and error bars denote the 95% confidence interval across seeds. Despite their much larger shifts in SF reliance, the LSF-biased models show only modest robustness gains. The extreme-LSF model does not outperform any of the NG models. The mixed-sigma LSF model performs better, but only reaches the level of the LO-guided model, even though the LO-guided model shows only a minimal shift in the LSF range. This dissociation indicates that increased LSF bias alone cannot account for the robustness advantage conferred by neural guidance.

**c.** Scatter plot comparing SF reliance shifts (from a.) and adversarial robustness gains (from b.) across NG models and LSF-biased models, similar to that shown in Fig. 2b. The LSF-biased models are shaped as triangles for visibility. The dissociation between SF reliance shifts and robustness gains is readily visible: NG models cluster along the y-axis with relatively small LSF shifts but a wide range of hierarchical robustness gains, whereas the LSF-biased models sit far to the right with substantially larger LSF shifts yet smaller robustness gains.

## Fig. S3

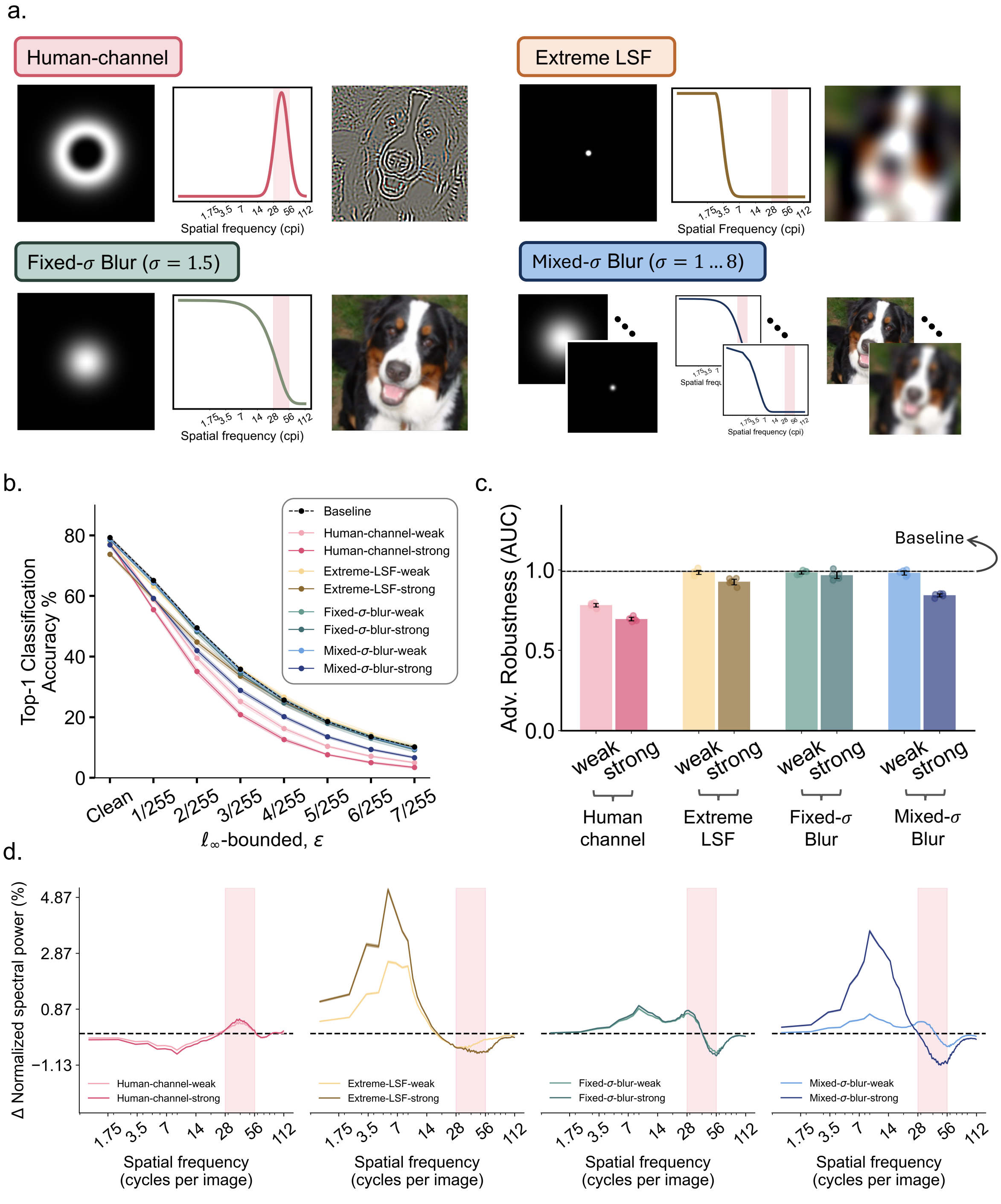

**Filtering-based spatial frequency (SF) biasing does not cleanly manipulate SF reliance and does not improve adversarial robustness.** Filtering-based methods to manipulate spatial

frequencies in natural images are more commonly used in the field [16,40]. However, we used the phase-scrambling-based method in the main experiments because we found that filtering-based biasing does not produce clean manipulations of SF reliance. We illustrate this here by repeating the same experiments shown in Fig. 3 using filtering-based SF biasing methods.

**a.** We first show the visualization of the SF filters and corresponding filtered images used to impose filtering-based SF biases during training. Similar to Fig. 3a, for each condition, we show the 2D frequency-domain filter mask (left), the corresponding 1D magnitude response plotted as a function of SF in cycles per image (cpi) on a logarithmic scale (middle), and example filtered images (right). The four conditions are: 1. the human-channel bandpass filter; 2. an extreme low spatial frequency (LSF) low-pass filter preserving frequencies below 3.5 cpi; 3. a fixed-$\sigma$ blur condition using a Gaussian kernel with standard deviation $\sigma = 1.5$, following [11]; and 4. a mix-$\sigma$ blur condition implemented as a mixture of Gaussian kernels with different $\sigma$ values, following [16]. We illustrate two example kernels here, $\sigma = 1.0$ and $\sigma = 8.0$. The human channel is highlighted by the pink shaded region in all 1D magnitude response plots. **b**. Adversarial robustness performance of models trained under each filtering-based SF bias condition. In all conditions except for baseline, we again included weak and strong SF bias variants by varying the proportion of filtered images during training. We used the same proportions (30% for weak bias and 50% for strong bias) as in the phase-scrambled-based experiments. All other training details are also kept identical. Top-1 classification accuracy under $\ell_\infty$-bounded PGD adversarial attacks [37] is shown across a range of perturbation bounds $\epsilon$. Shaded regions indicate bootstrapped 95% confidence intervals (CI) estimated from five independent models each trained with a different random seed. **c**. Area under the robustness curve (AUC) summarizing adversarial performance for each condition shown in **b**., with higher AUC indicating greater robustness. Error bars denote bootstrapped 95% CI.

d. SF reliance profiles for all filtering-based SF-biased models, shown as differences relative to the baseline model, similar to Fig. 2a. From left to right, conditions are human-channel-biased, extreme-LSF-biased, fixed-$\sigma$-biased, and mix-$\sigma$-biased. Positive values indicate greater reliance at a given SF relative to the baseline model. Weak and strong variants within each condition are plotted together. The human channel is highlighted by the pink shaded region.

Our results show that filter-based biasing did not elicit clean, selective SF biases. For example, the human-channel condition showed only a relatively small shift in reliance, whereas the extreme-LSF condition showed its most prominent increase in reliance in additional octaves beyond the targeted range. Furthermore, none of the filtering-based models showed improved adversarial robustness relative to the baseline, which further supports the conclusion that simply biasing models towards selected SF ranges is not sufficient to explain the robustness advantage associated with neural guidance. One possible reason is that filtered images only reduce the availability of untargeted SF cues, that is, SF ranges outside the targeted band, rather than actively discouraging reliance on them. DCNNs are well known to exploit any predictive cues that remain available [57], and thus filtering-based methods may not impose a strong enough bias. By contrast, the phase-scrambling-based method not only preserves the targeted band but also actively disrupts the untargeted bands, thereby imposing a stronger bias. Therefore, we used the phase-scrambling-based SF biasing method in the main experiments.